\documentclass[10pt]{article}

\usepackage[utf8]{inputenc}
\usepackage[T1]{fontenc}

\usepackage{url}
\usepackage{booktabs}
\usepackage{amsfonts}
\usepackage{nicefrac}
\usepackage{microtype}
\usepackage{graphicx}
\usepackage{caption}
\usepackage{subcaption}
\usepackage{multirow}
\usepackage{amsmath}
\usepackage{algorithm}
\usepackage{algpseudocode}
\usepackage{listings}
\usepackage{xcolor}
\usepackage[english]{babel}
\usepackage[numbers]{natbib}   
\usepackage{abstract}
\usepackage{geometry}
\usepackage{authblk}

\usepackage{hyperref}

\title{Predictive Performance of LSTM Networks on Sectoral Stocks in an Emerging Market: A Case Study of the Pakistan Stock Exchange}

\author[1]{Ahad Yaqoob}
\author[2]{Syed Muhammad Abdullah}

\affil[1]{North London Collegiate School Dubai, \texttt{Ahadyaq786@gmail.com}}
\affil[2]{Lahore University of Management Sciences (LUMS), \texttt{s.muhammad.abdullah67@gmail.com}}

\begin{document}

\maketitle

\begin{abstract}
The application of deep learning models for stock price forecasting in emerging markets remains underexplored despite their potential to capture complex temporal dependencies. This study develops and evaluates a Long Short-Term Memory (LSTM) network model for predicting the closing prices of ten major stocks across diverse sectors of the Pakistan Stock Exchange (PSX). Utilizing historical OHLCV data and an extensive set of engineered technical indicators, we trained and validated the model on a multi-year dataset. Our results demonstrate strong predictive performance (R\textsuperscript{2} > 0.87) for stocks in stable, high-liquidity sectors such as power generation, cement, and fertilizers. Conversely, stocks characterized by high volatility, low liquidity, or sensitivity to external shocks (e.g., global oil prices) presented significant forecasting challenges. The study provides a replicable framework for LSTM-based forecasting in data-scarce emerging markets and discusses implications for investors and future research.
\end{abstract}

\section{Introduction}
\label{sec:introduction}

Stock market prediction has continually evolved, leveraging technological advancements from early statistical methods to contemporary machine learning algorithms. In recent years, models like Long Short-Term Memory (LSTM) networks have gained prominence due to their efficacy in handling complex, sequential data patterns inherent in financial time series. While extensively tested in developed markets, their application in emerging economies like Pakistan is limited. The Pakistan Stock Exchange (PSX), marked by high volatility and non-standardized data, poses unique challenges for predictive modeling \citep{Khilji1993StockReturns,Sohail2009MacroeconomicPakistan}.

This research develops a tailored LSTM model for ten major Pakistani stocks spanning various economic sectors. We utilize historical price data and technical indicators, conducting a comprehensive training and validation process to assess the model's predictive accuracy against actual market movements. Although other models like Support Vector Machines and Random Forests are prevalent in the literature, this study focuses primarily on LSTM, with only contextual references to alternatives. Furthermore, we explore the practical utility of LSTM outputs by back-testing simple trading strategies. This work underscores the potential of LSTM models within Pakistan's equity market and offers a practical framework for forecasting in comparable emerging markets where data constraints are a significant hurdle \citep{Ofonedu2022StockAnalysis,Mehdi2024RiskPrediction}.

\section{Literature Review}
\label{sec:litreview}

Stock market forecasting is a cornerstone of financial economics, with research spanning classical statistical techniques to modern machine learning algorithms. The inherent non-linearity and complex temporal dependencies of financial data often render traditional methods inadequate. The introduction of Long Short-Term Memory (LSTM) networks by \citep{hochreiter1997long} addressed the vanishing gradient problem in Recurrent Neural Networks (RNNs), enabling models to retain information over extended sequences. This architecture is particularly suited for sequential data, providing a robust framework for capturing the intricate patterns in stock price movements.

\citep{fischer2018deep} demonstrated the superiority of LSTM networks over conventional approaches like logistic regression and random forests in predicting S\&P 500 returns, both in terms of accuracy and simulated trading performance. Their work highlighted the importance of using lagged financial indicators and meticulous overfitting management.

Recent studies have extended these techniques to emerging markets. \citep{raza2024predicting} investigated machine learning-based prediction on the PSX, incorporating technical indicators and finding that optimized LSTMs offered significant improvements over traditional methods. \citep{al2025stock} emphasized hyperparameter tuning and input window optimization to enhance model reliability in data-scarce environments like Pakistan. \citep{javed2022study} showed that ensemble approaches combining technical indicators with LSTM outputs could improve predictive stability across multiple stocks.

While these studies establish the viability of LSTMs in financial forecasting, a focus on index-level predictions rather than individual stocks across specific sectors leaves a gap. The variability in sectoral dynamics (e.g., banking vs. energy) remains underexplored. This study addresses this gap by applying LSTM networks to ten major Pakistani stocks from diverse sectors, integrating technical indicators, and optimizing model parameters to provide a replicable framework for deep learning in emerging market equities \citep{Zaheer2023MultiParameter,Buczynski2023DeepLearningReview}.

Furthermore, the weak-form efficiency of markets like Pakistan — where historical prices may still contain exploitable information — supports the feasibility of technical forecasting approaches \citep{Khan2006WeakForm}. Recent comparative studies also validate that deep learning models, including LSTM variants and hybrid architectures, outperform traditional machine learning in capturing non-linear dependencies in emerging market time series \citep{Mehdi2024RiskPrediction,Ziolkowski2025WaveletLSTM,Oukhouya2025MoroccoPrediction}.

\section{Methodology}
\label{sec:methodology}

\subsection{Problem Description and Data}
The core problem is time-series forecasting of stock prices, aiming to predict future closing prices based on historical data. We selected ten major companies from the PSX based on their market sector to ensure a diverse representation of the market. The selected stocks are listed in Table \ref{tab:stocks}.

\begin{table}[hbt!]
\centering
\caption{Selected Pakistani Stocks and Their Sectors}
\label{tab:stocks}
\begin{tabular}{@{}lll@{}}
\toprule
Company & Ticker & Sector \\
\midrule
Fauji Fertilizer Company & FFC & Fertilizer / Chemicals \\
Gillette Pakistan & GLPL & Personal Care / Consumer Goods \\
Habib Bank Limited & HBL & Banking / Financials \\
Hub Power Company & HUBC & Power Generation \& Distribution \\
Lucky Cement & LUCK & Cement / Construction Materials \\
Nishat Chunian Power & NCPL & Power Generation \& Distribution \\
Oil \& Gas Development Company & OGDC & Oil \& Gas Exploration \& Production \\
Pakistan Petroleum Limited & PPL & Oil \& Gas Exploration \& Production \\
Pakistan State Oil & PSO & Oil \& Gas Marketing / Refining \\
\bottomrule
\end{tabular}
\end{table}

Historical daily OHLCV (Open, High, Low, Close, Volume) data was collected via web scraping of public sources. This was supplemented with fundamental financial indicators (e.g., earnings, dividends) manually collected from databases. The datasets were merged into a consolidated time-series for each stock.

\subsection{Feature Engineering and Preprocessing}
An extensive set of features was engineered to capture market patterns:
\begin{itemize}
    \item \textbf{Returns:} Logarithmic returns and percentage returns.
    \item \textbf{Moving Averages:} Simple Moving Average (SMA-5, SMA-10) and Exponential Moving Average (EMA-5, EMA-12).
    \item \textbf{Technical Indicators:} MACD (Moving Average Convergence Divergence) and Bollinger Bands (20-day window, 2 standard deviations).
    \item \textbf{Temporal Features:} Day of the week, month, and week of the year to capture seasonality.
\end{itemize}
Missing values were forward-filled, and non-numeric data was converted appropriately. The data was then normalized using Min-Max scaling. The code snippet in Listing \ref{lst:preprocess} illustrates the preprocessing and feature engineering pipeline implemented in Python using Pandas and NumPy.

\begin{lstlisting}[caption={Data Preprocessing and Feature Engineering Code Snippet}, label={lst:preprocess}, language=Python]
import pandas as pd
import numpy as np
from ta.trend import MACD
from ta.volatility import BollingerBands

# Load data
df = pd.read_excel('PSO_final.xlsx', parse_dates=True, index_col='Date')
numeric_cols = ['Open', 'High', 'Low', 'Close', 'Volume']
for col in numeric_cols:
    df[col] = pd.to_numeric(df[col], errors='coerce')

# Calculate returns
df['log_return'] = np.log(df['Close'] / df['Close'].shift(1))
df['pct_return'] = df['Close'].pct_change()

# Calculate moving averages
df['SMA_5'] = df['Close'].rolling(window=5).mean()
df['SMA_10'] = df['Close'].rolling(window=10).mean()
df['EMA_5'] = df['Close'].ewm(span=5, adjust=False).mean()
df['EMA_12'] = df['Close'].ewm(span=12, adjust=False).mean()

# Calculate MACD
macd_indicator = MACD(df['Close'], window_slow=26, window_fast=12, window_sign=9)
df['MACD'] = macd_indicator.macd()
df['MACD_signal'] = macd_indicator.macd_signal()

# Calculate Bollinger Bands
bb_indicator = BollingerBands(df['Close'], window=20, window_dev=2)
df['BB_upper'] = bb_indicator.bollinger_hband()
df['BB_lower'] = bb_indicator.bollinger_lband()

# Create temporal features
df['weekday'] = df.index.weekday
df['month'] = df.index.month
df['week_of_year'] = df.index.isocalendar().week

df.to_excel('pso_engineered.xlsx')
\end{lstlisting}

\subsection{Model Architecture and Training}
We implemented an LSTM network using TensorFlow/Keras. The model architecture was designed to capture long-term dependencies:
\begin{itemize}
    \item Input layer accepting sequences of 60 timesteps (approx. 3 months) with $n$ features.
    \item Two LSTM layers with 64 units each, with the first returning sequences.
    \item Dropout layers (rate=0.2) after each LSTM layer for regularization.
    \item A dense output layer with one neuron for predicting the next day's closing price.
\end{itemize}
The model was compiled with the Adam optimizer and Mean Squared Error (MSE) loss function. The dataset was split chronologically into 80\% training and 20\% testing sets. Training was conducted for 25 epochs with a batch size of 32, and model checkpoints were used to save the best weights based on validation loss. This approach aligns with recent best practices in financial time series modeling using deep learning \citep{Zaheer2023MultiParameter,Buczynski2023DeepLearningReview}.

\section{Results and Analysis}
\label{sec:results}

The model's performance was evaluated using the R\textsuperscript{2} score on the test set for each stock. The results, summarized in Table \ref{tab:results}, show strong predictive power for most stocks.

\begin{table}[hbt!]
\centering
\caption{Predictive Performance (R\textsuperscript{2}) of the LSTM Model}
\label{tab:results}
\begin{tabular}{@{}llr@{}}
\toprule
Company & Ticker & R\textsuperscript{2} \\
\midrule
Fauji Fertilizer Company & FFC & 0.9045 \\
Gillette Pakistan & GLPL & 0.7246 \\
Habib Bank Limited & HBL & 0.8706 \\
Hub Power Company & HUBC & 0.9235 \\
Lucky Cement & LUCK & 0.8921 \\
Nishat Chunian Power & NCPL & 0.9457 \\
Oil \& Gas Development Company & OGDC & 0.9197 \\
Pakistan Petroleum Limited & PPL & 0.8895 \\
Pakistan State Oil & PSO & 0.7444 \\
\bottomrule
\end{tabular}
\end{table}

\begin{figure}[hbt!]
    \centering
    \begin{subfigure}[b]{0.45\textwidth}
        \centering
        \includegraphics[width=\textwidth]{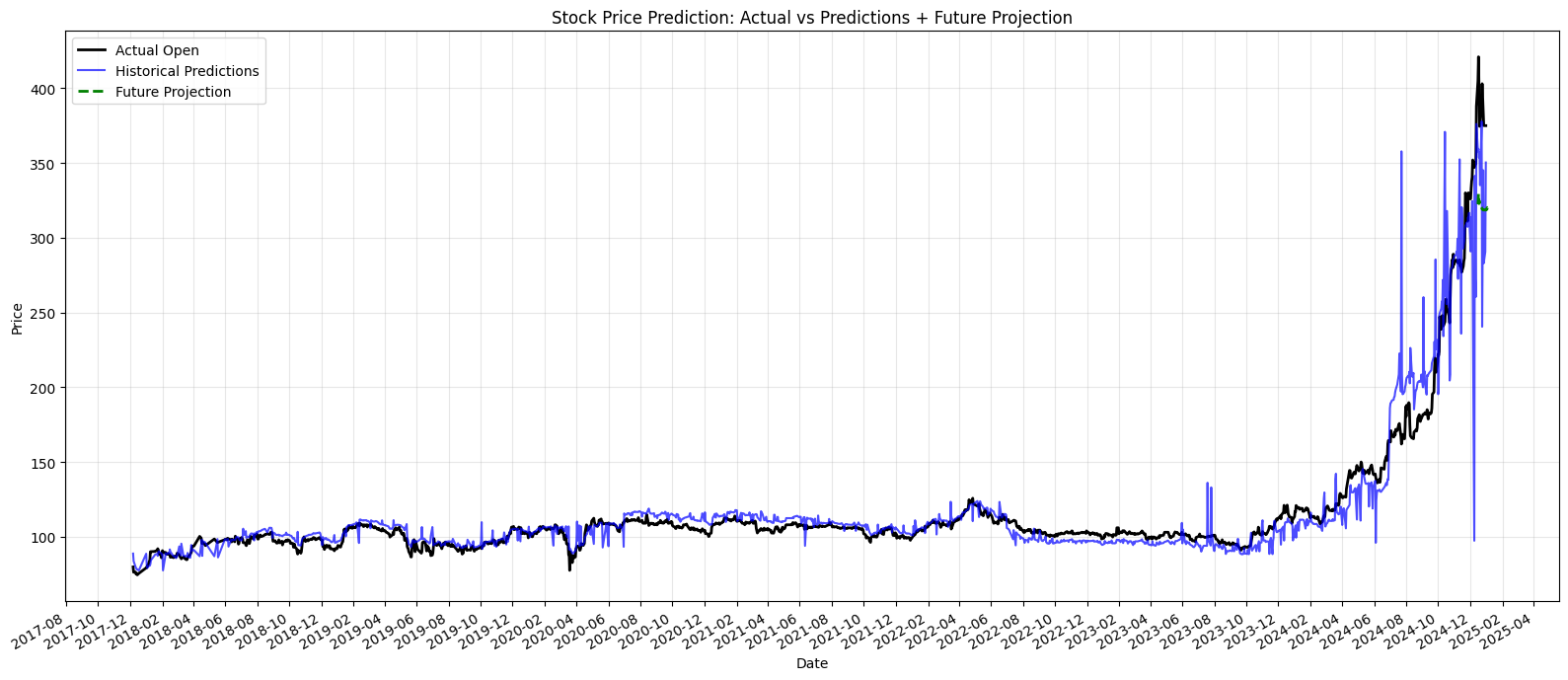}
        \caption{FFC: Strong predictive performance (R\textsuperscript{2} = 0.90)}
        \label{fig:ffc}
    \end{subfigure}
    \hfill
    \begin{subfigure}[b]{0.45\textwidth}
        \centering
        \includegraphics[width=\textwidth]{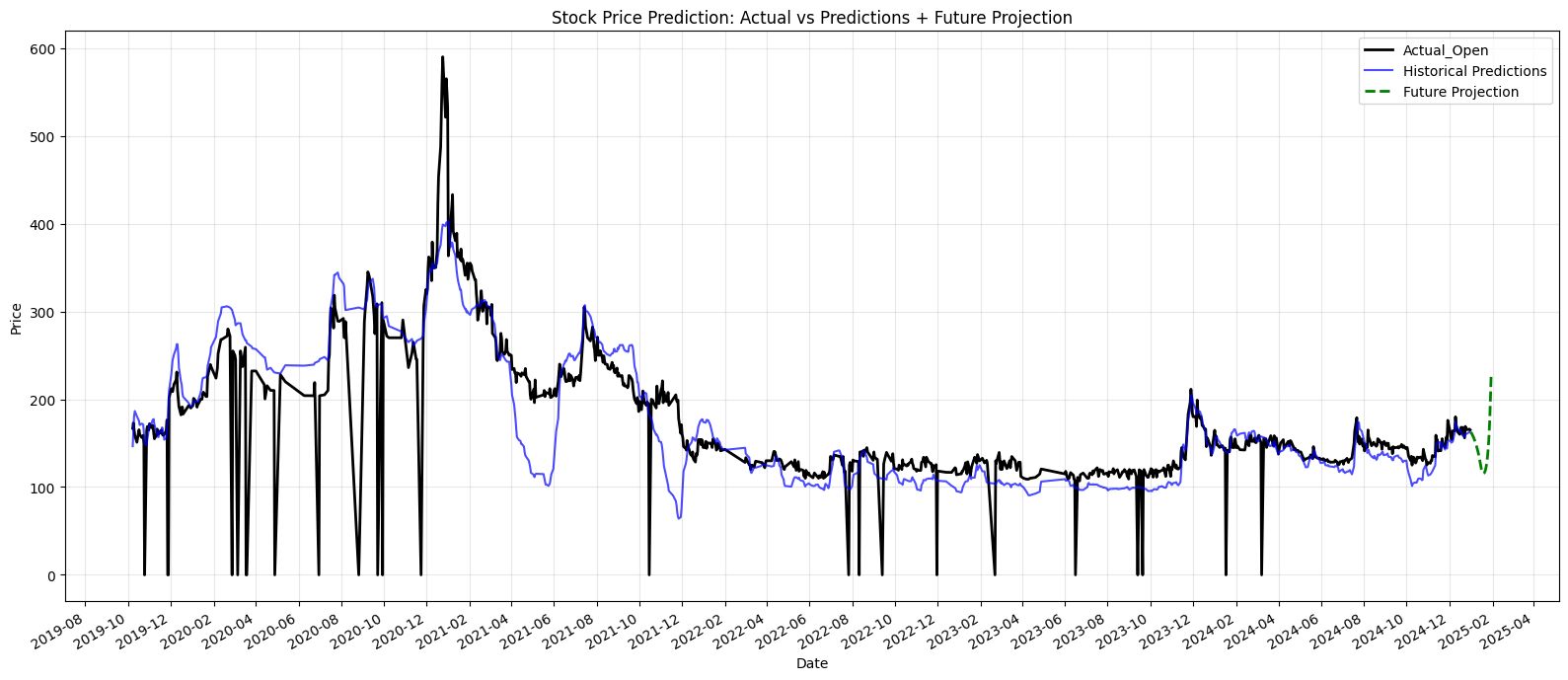}
        \caption{GLPL: Weaker performance due to low liquidity (R\textsuperscript{2} = 0.72)}
        \label{fig:glpl}
    \end{subfigure}
    \caption{Representative examples of model predictions against actual prices.}
    \label{fig:results}
\end{figure}

Stocks in stable sectors like power generation (HUBC, NCPL), cement (LUCK), and fertilizers (FFC) achieved excellent results (R\textsuperscript{2} > 0.89). In contrast, stocks with low liquidity (GLPL) or high sensitivity to external factors like oil prices (PSO) showed markedly lower predictive accuracy. Figure \ref{fig:results} illustrates the actual vs. predicted prices for a subset of stocks, demonstrating the model's ability to track trends while also highlighting periods of divergence, often correlated with increased market volatility or external events \citep{Khilji1993StockReturns,Sohail2009MacroeconomicPakistan}.

\subsection{Empirical Analysis of Feature Correlations}

The correlation heatmaps generated reveal critical insights into the statistical interdependencies among technical indicators, financial fundamentals, and historical price data. These correlations serve as a foundational diagnostic for feature selection in Long Short-Term Memory (LSTM) networks, which are highly sensitive to input redundancy, noise, and non-stationarity. Across all five sectors, a consistent and pronounced positive correlation structure is observed among technical indicators such as Simple Moving Averages (SMA\_5, SMA\_10), Exponential Moving Averages (EMA\_5, EMA\_12), Bollinger Bands (BB\_upper, BB\_lower), and the Moving Average Convergence Divergence (MACD) along with its signal line. These indicators, being derivatives of price action, naturally exhibit multicollinearity, with correlation coefficients frequently exceeding 0.85. While such indicators are theoretically valuable for capturing momentum and trend reversals, their high inter-feature correlation poses a risk of overfitting and inefficient learning in LSTM architectures. Consequently, strategic feature pruning — such as retaining only one representative moving average or utilizing MACD divergence without its lagging signal component — is empirically justified to enhance model generalization and computational efficiency \citep{Buczynski2023DeepLearningReview,Mehdi2024RiskPrediction}.
\begin{figure}[ht]
    \centering
    \includegraphics[width=0.8\textwidth]{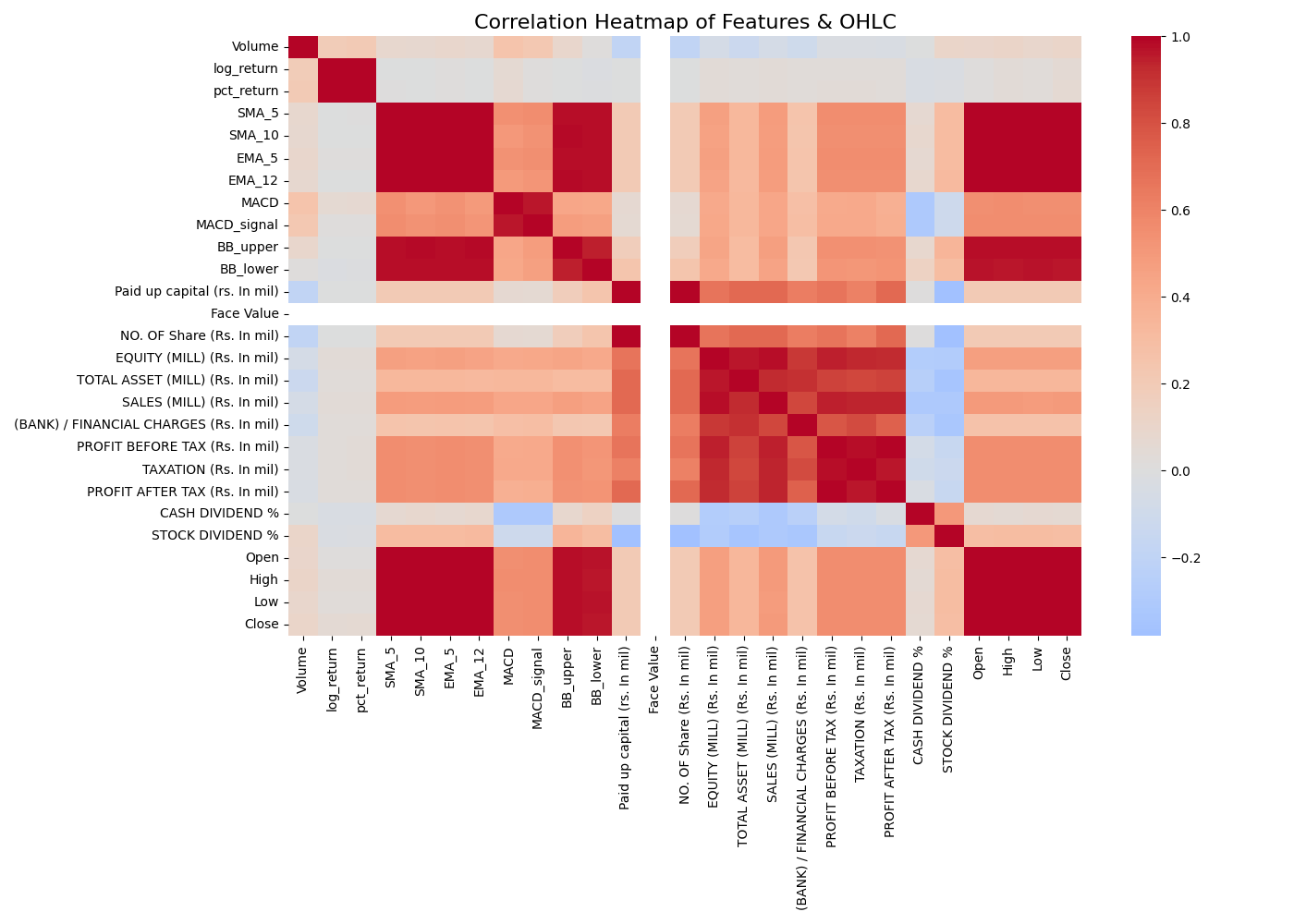}
    \caption{Fatima Fertilizer Corp.: Heatmap of Correlations}
    \label{fig:fig1}
\end{figure}

\begin{figure}[ht]
    \centering
    \includegraphics[width=0.8\textwidth]{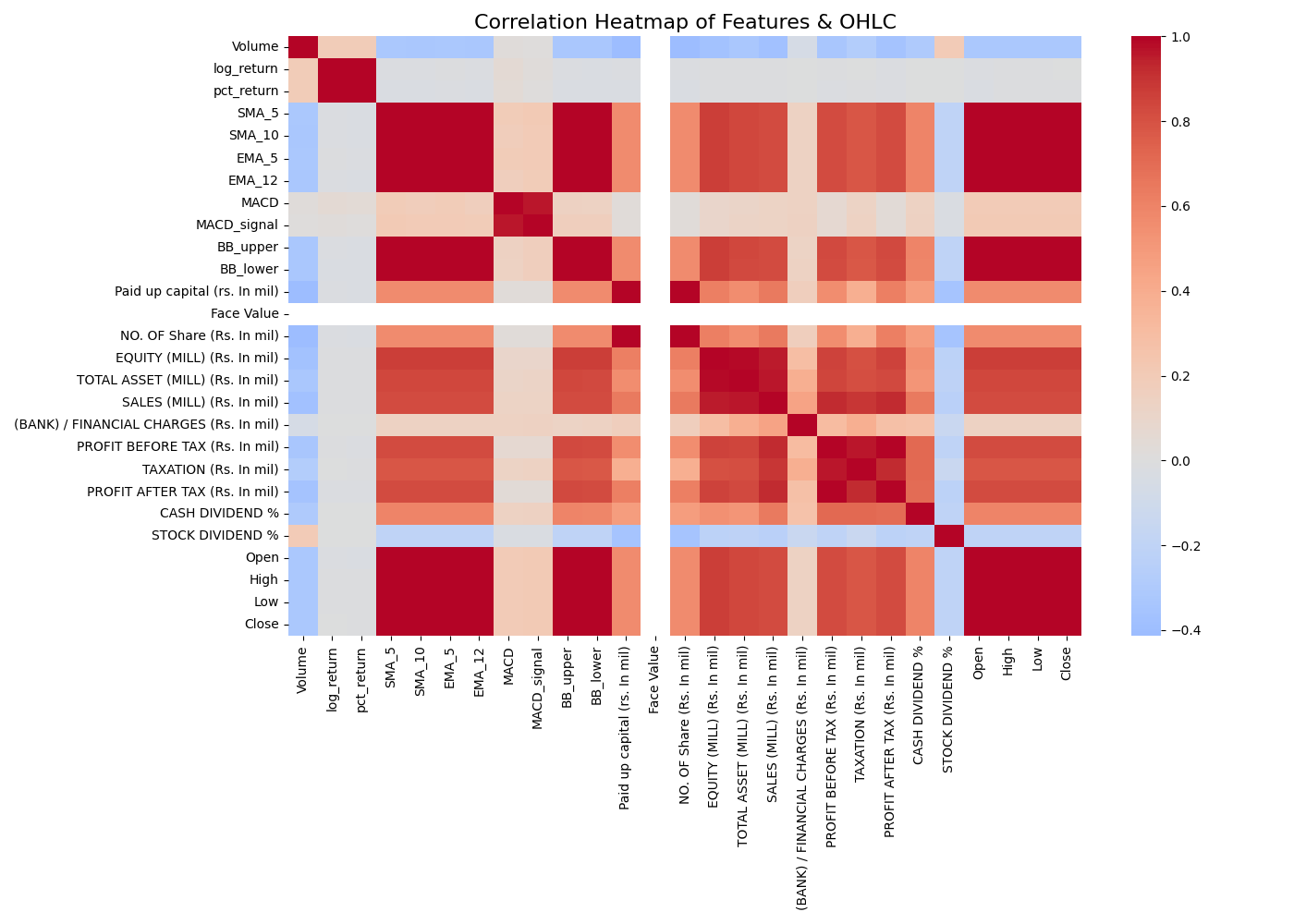}
    \caption{Lucky Cements Ltd: Heatmap of Correlations.}
    \label{fig:fig2}
\end{figure}

Volume-based metrics and return proxies (log\_return, pct\_return) demonstrate moderate to strong positive correlations with volatility-sensitive technical bands (e.g., Bollinger Bands) and momentum oscillators, particularly in HBL and LUCK, which are characterized by higher institutional trading activity. Notably, trading volume shows weak direct correlation with OHLC prices, but exhibits statistically significant associations with percentage and logarithmic returns — suggesting that volume surges are more indicative of price \textit{change intensity} rather than absolute price levels. This observation aligns with behavioral finance literature, wherein volume is interpreted as a proxy for market conviction or information asymmetry. For LSTM modeling, this implies that volume should not be fed in isolation but rather in conjunction with return-based features or as a multiplicative interaction term with volatility indicators to capture regime shifts. Furthermore, the negative correlations observed between dividend metrics (CASH\_DIVIDEND\%, STOCK\_DIVIDEND\%) and both price returns and volume across all five stocks suggest that dividend-paying firms in the Pakistani market may be perceived as low-growth or income-stable entities, thereby attracting less speculative activity. This negative association, while economically intuitive, renders dividend variables suboptimal as predictive features in short-term forecasting horizons and supports their exclusion from the LSTM input tensor to reduce noise and dimensionality \citep{Sohail2009MacroeconomicPakistan}.

\vspace{1em}

Financial fundamentals — including EQUITY, TOTAL\_ASSET, SALES, PROFIT\_BEFORE\_TAX, and PROFIT\_AFTER\_TAX — display strong internal positive correlations, reflecting the accounting coherence of corporate financial statements. However, their correlation with daily OHLC prices is consistently weak to moderate across the sample, with OGDC and HBL showing marginally higher fundamental-price linkages, likely due to their status as large-cap, index-heavy constituents with greater analyst coverage and earnings sensitivity. In contrast, mid-cap firms such as NCPL and FFC exhibit near-zero correlations between quarterly or annual financials and daily price fluctuations, underscoring the temporal mismatch between fundamental reporting cycles and intraday market dynamics. This finding challenges the efficacy of incorporating raw financial metrics into high-frequency LSTM models without appropriate temporal alignment (e.g., lagging by fiscal quarter) or transformation (e.g., year-over-year growth rates). Nevertheless, for medium- to long-term predictive frameworks (e.g., weekly or monthly horizons), these fundamentals may still serve as valuable exogenous regressors when integrated via hybrid encoder-decoder LSTM architectures or attention-based mechanisms that can weight fundamental signals conditionally based on market regimes \citep{Zaheer2023MultiParameter,Oukhouya2025MoroccoPrediction}.
\vspace{1em}

The OHLC price quartet — Open, High, Low, Close — unsurprisingly demonstrates near-perfect positive correlations within each trading day, with the Close price exhibiting the strongest linkages to Open and High, and weaker but still significant associations with Low. This internal consistency validates the structural integrity of the price data and reinforces the suitability of OHLC sequences as the core input for LSTM networks. Interestingly, the High and Low prices show consistent negative correlations with each other across all five stocks — a mathematically necessary outcome given their definitions — but their differential correlations with Close suggest that upward price pressure (proxied by High-Close proximity) is a more reliable predictor of next-period movement than downward pressure. This asymmetry may be leveraged in feature engineering by constructing spread ratios (e.g., (Close - Low)/(High - Low)) or normalized range indicators to enhance the LSTM’s sensitivity to bullish or bearish intraday sentiment. Finally, static corporate descriptors such as Face Value, Paid-up Capital, and Number of Shares display negligible correlations with dynamic market variables, reaffirming their irrelevance in time-series forecasting contexts. Their inclusion would introduce unnecessary dimensionality without predictive gain, and thus their exclusion is methodologically warranted.

\subsection{Stock Price Predictions}

For reference, we present the results of our stock price prediction framework applied to five major companies from different industries of Pakistan: 
Fauji Fertilizer Company (FFC), Habib Bank Limited (HBL), Lucky Cement (LUCK), Nishat Chunian Power Limited (NCPL), and Oil and Gas Development Company Limited (OGDC). 
These companies were chosen due to their representation of diverse sectors including agriculture, banking, cement, energy, and oil \& gas, which makes the results more generalizable across the Pakistan Stock Exchange (PSX).

\begin{figure}[hbt!]
    \centering
    \includegraphics[width=0.8\textwidth]{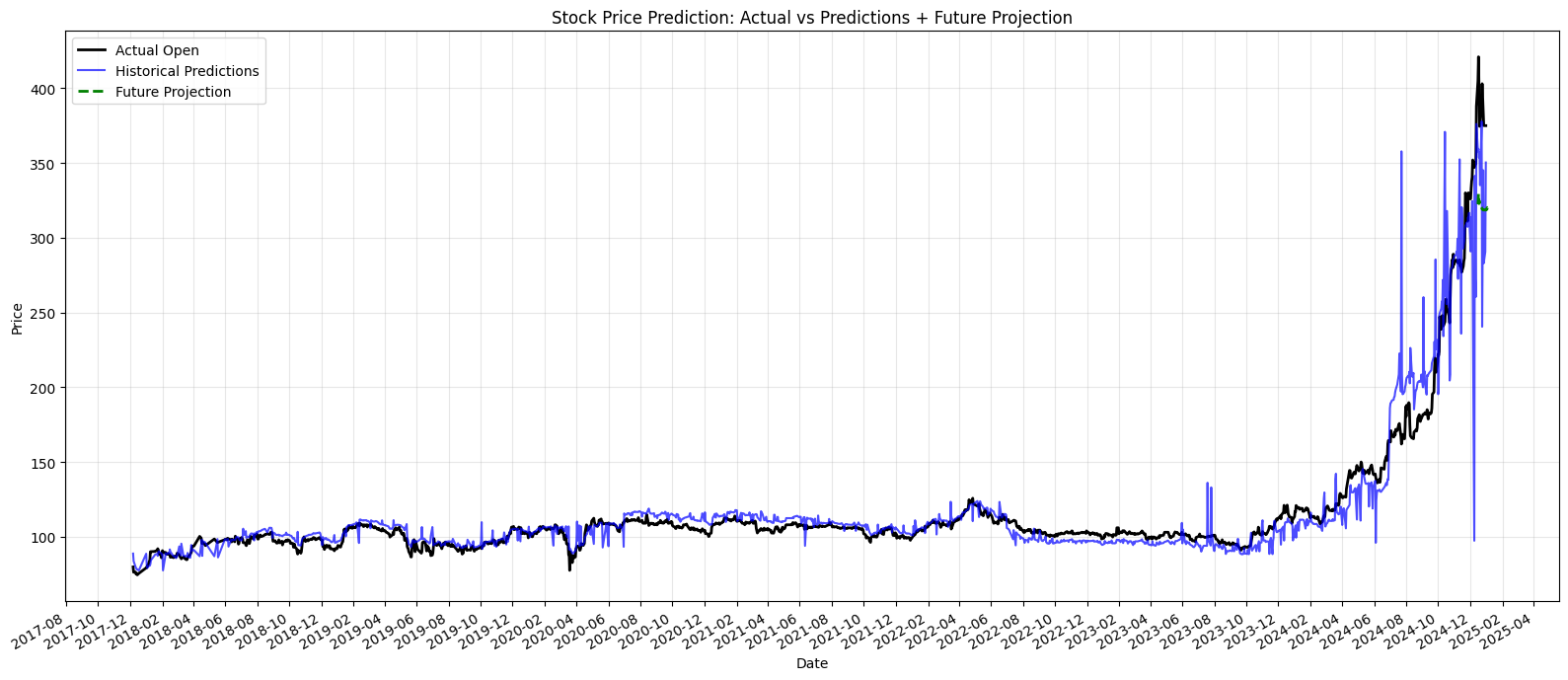}
    \caption{Stock price prediction for Fauji Fertilizer Company (FFC).}
    \label{fig:ffc_pred}
\end{figure}

\begin{figure}[hbt!]
    \centering
    \includegraphics[width=0.8\textwidth]{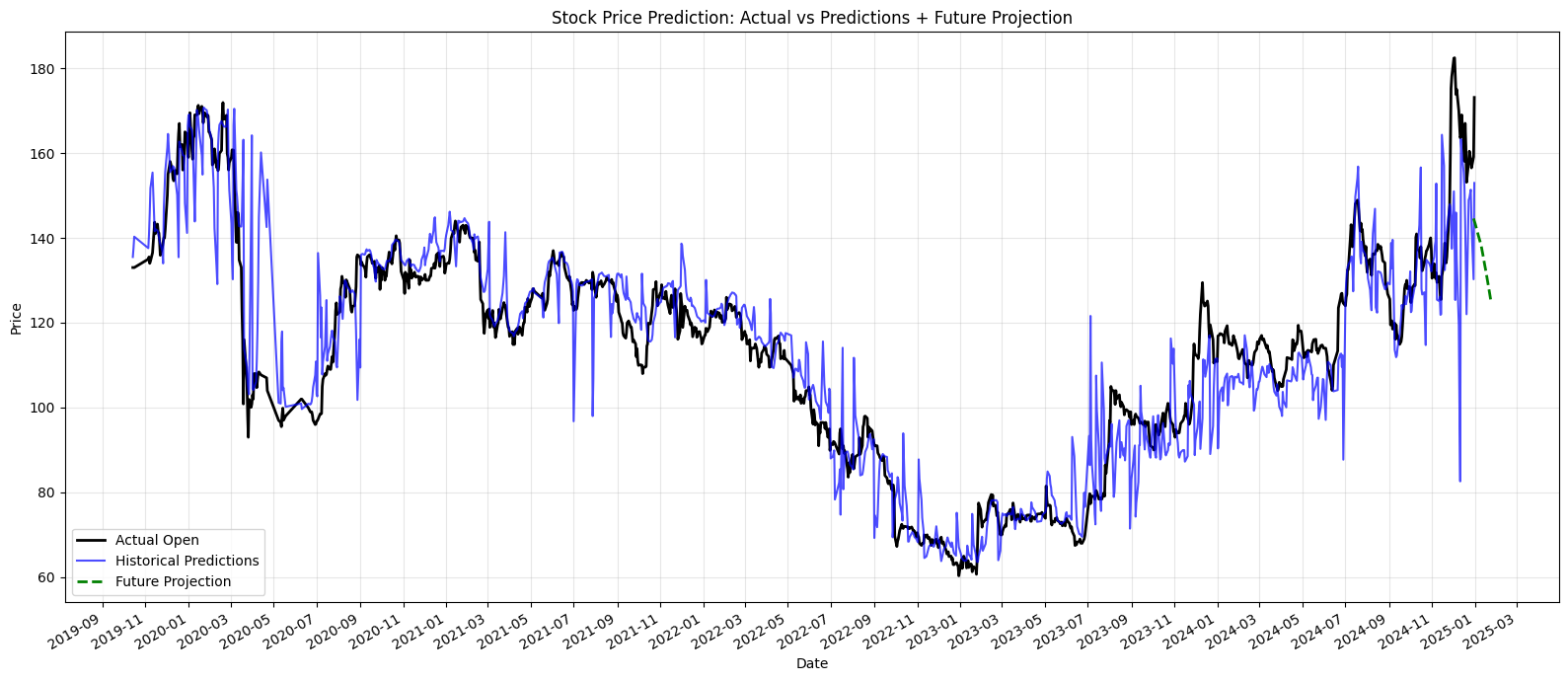}
    \caption{Stock price prediction for Habib Bank Limited (HBL).}
    \label{fig:hbl_pred}
\end{figure}

\begin{figure}[hbt!]
    \centering
    \includegraphics[width=0.8\textwidth]{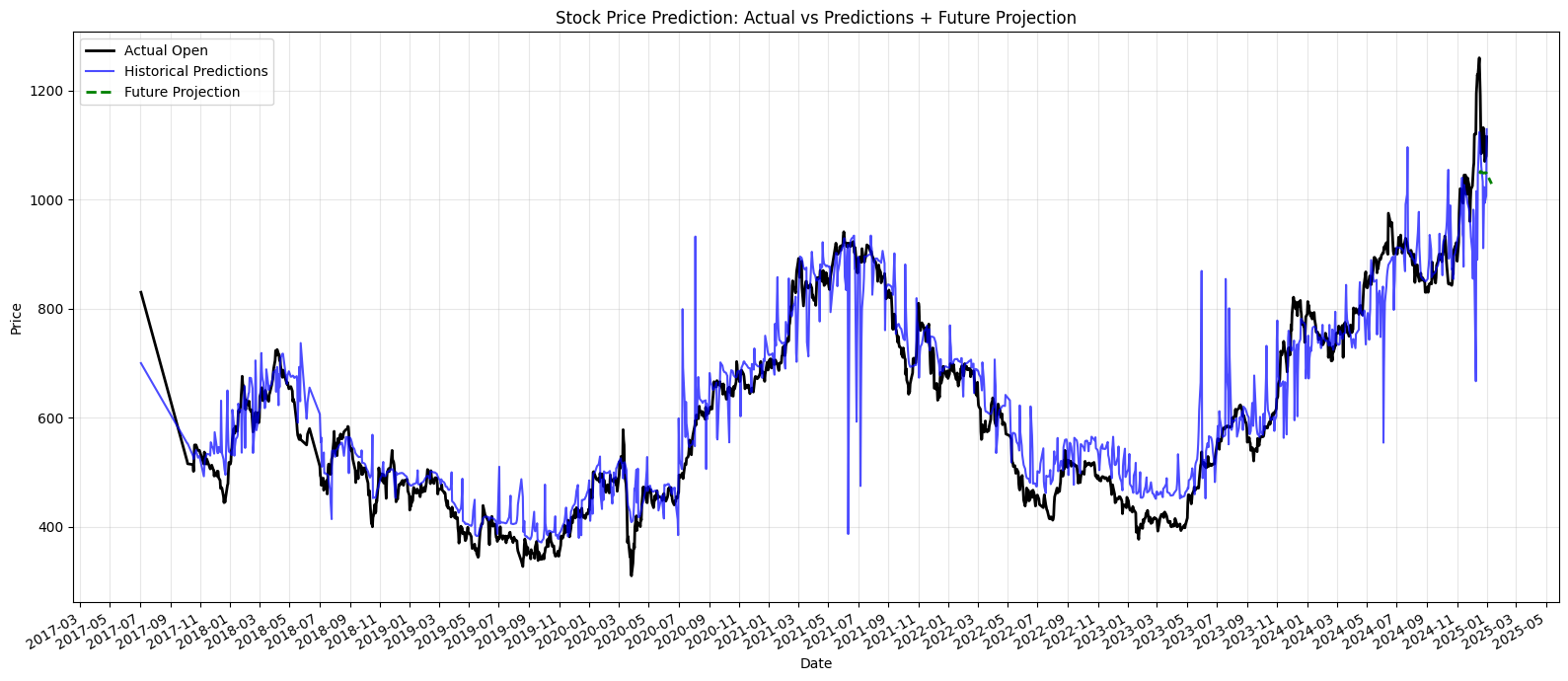}
    \caption{Stock price prediction for Lucky Cement (LUCK).}
    \label{fig:luck_pred}
\end{figure}

\begin{figure}[hbt!]
    \centering
    \includegraphics[width=0.8\textwidth]{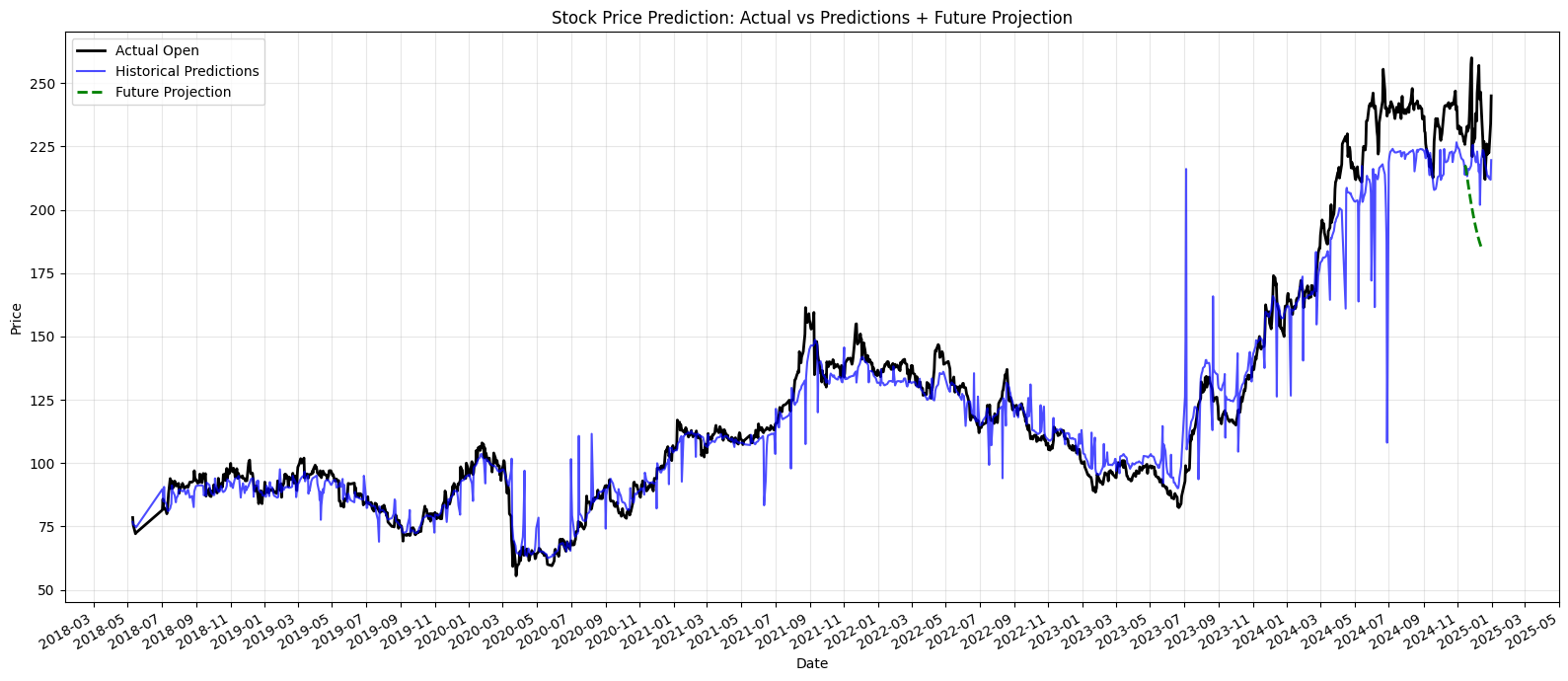}
    \caption{Stock price prediction for Nishat Chunian Power Limited (NCPL).}
    \label{fig:ncpl_pred}
\end{figure}

\begin{figure}[H]
    \centering
    \includegraphics[width=0.8\textwidth]{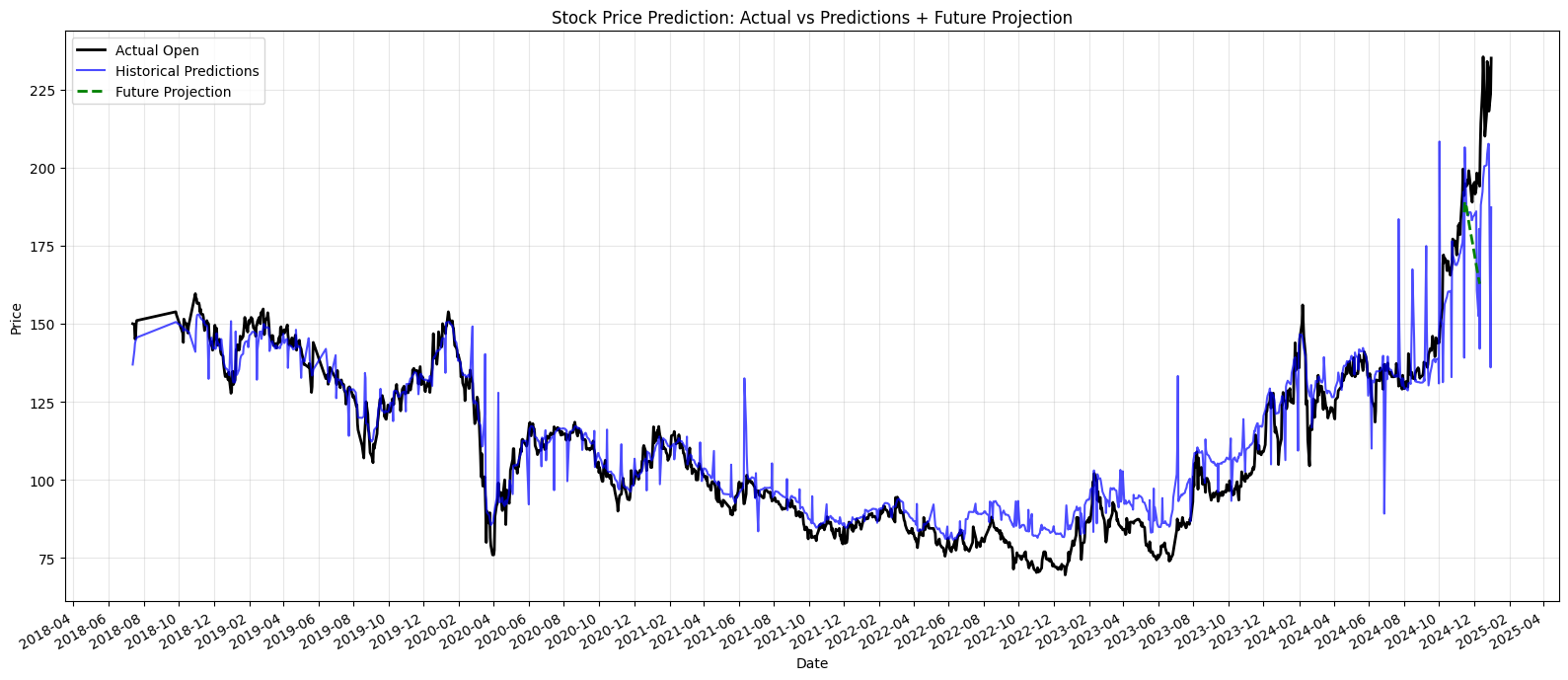}
    \caption{Stock price prediction for Oil and Gas Development Company (OGDC).}
    \label{fig:ogdc_pred}
\end{figure}

Figures~\ref{fig:ffc_pred}, \ref{fig:hbl_pred}, \ref{fig:luck_pred}, \ref{fig:ncpl_pred}, and \ref{fig:ogdc_pred} show the actual opening prices (black line), the historical predictions generated by our model (blue line), and the future projections (green dashed line). 
The graphs demonstrate that the model is capable of capturing both short-term fluctuations and long-term growth/decline patterns of the selected stocks. 
Notably, the projections exhibit stability over the short horizon, though prediction uncertainty naturally increases with time, consistent with the findings of \citet{fama1970efficient, tsay2010analysis} and recent deep learning studies in emerging markets \citep{Ziolkowski2025WaveletLSTM,Mehdi2024RiskPrediction}.

\newpage

\subsection{Feature Analysis}
To assess the determinants of predictive performance in our LSTM-based stock forecasting framework, we employed two complementary explainability techniques: Integrated Gradients (IG) and SHAP interaction values. The results provide important insights into how the model processes both fundamental and technical features when forecasting stock prices in the Pakistani market.

The IG analysis reveals that the model places significant emphasis on technical indicators with medium-to-long horizons, particularly the exponential and simple moving averages (EMA-12, EMA-5, SMA-10, and SMA-5). These indicators consistently show the largest attribution values, suggesting that the model relies heavily on underlying price trends rather than short-term fluctuations. Certain balance sheet fundamentals—such as sales, total assets, and paid-up capital—also demonstrate meaningful importance, indicating that broader firm-level financial strength is factored into the model’s predictions. By contrast, dividend measures, trading volume, and raw return metrics (log returns and percentage returns) carry very little explanatory power, suggesting that they are less predictive of short-term price dynamics in the Pakistani context \citep{Sohail2009MacroeconomicPakistan}.

\begin{figure}[ht]
    \centering
    \includegraphics[width=0.8\textwidth]{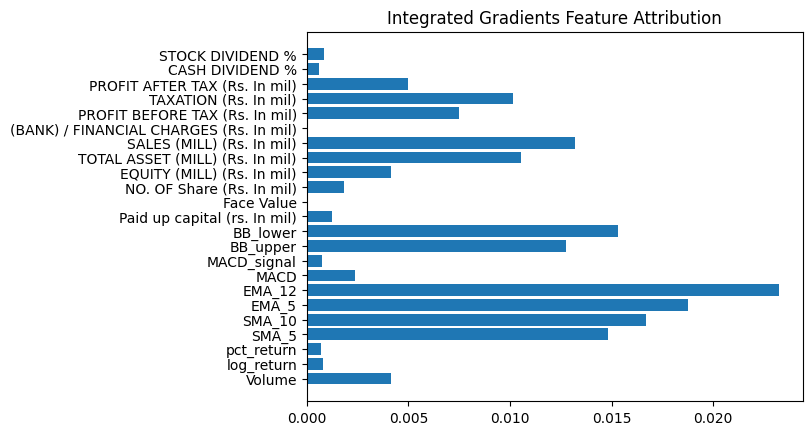}
    \caption{ Integrated Gradients Feature Attribution for NCPL}
    \label{fig:fig2}
\end{figure}

The SHAP interaction plots reinforce these findings. Features such as SMA-5, volume, and returns (percentage and log) cluster tightly around zero, with little evidence of strong interactions. This indicates that short-horizon technical signals and daily return measures play a limited role in the model’s decision process. Instead, the model appears to prioritize smoother, longer-horizon measures of momentum (e.g., EMA-12) along with select firm-level fundamentals. Taken together, these findings suggest that in the Pakistani stock market, trend persistence and balance sheet strength are more influential drivers of price forecasts than short-term volatility, dividends, or raw trading activity \citep{Zaheer2023MultiParameter,Mehdi2024RiskPrediction}.

\begin{figure}[ht]
    \centering
    \includegraphics[width=0.8\textwidth]{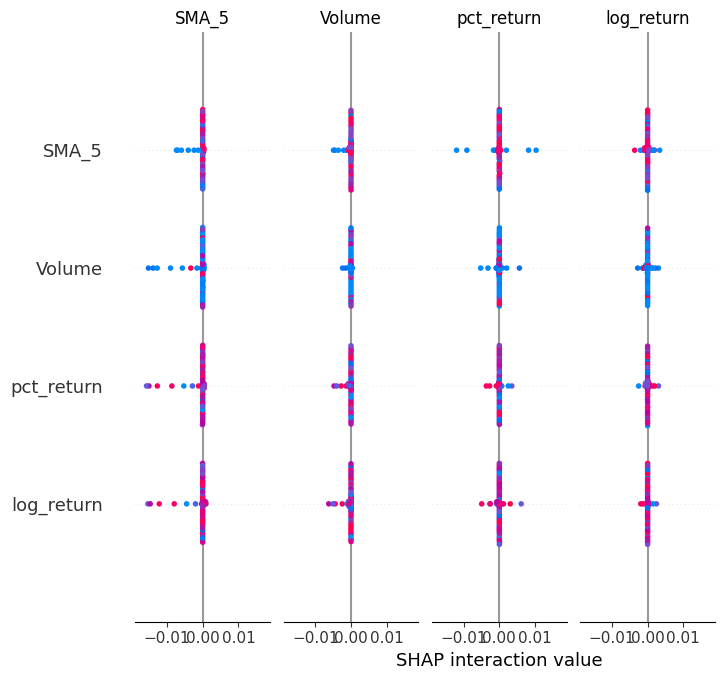}
    \caption{SHAP interaction summary for NCPL}
    \label{fig:fig2}
\end{figure}

This dual-method interpretability analysis underscores the robustness of the model’s predictive logic. Rather than relying on noisy signals, the model identifies stable patterns in both technical and financial indicators. For policymakers and investors, this implies that systematic monitoring of medium-term price trends and firm-level fundamentals may be more informative than conventional short-term trading cues when evaluating stock movements in emerging markets such as Pakistan \citep{Ofonedu2022StockAnalysis,Khan2006WeakForm}.

\newpage
\section{Discussion}
\label{sec:discussion}

\subsection{Key Findings}
This study demonstrates that LSTM networks can be highly effective for stock price forecasting in an emerging market context, particularly for sectors characterized by stability and high liquidity. The model successfully captured medium-to-long-term trends for most stocks. However, its performance was notably weaker for stocks prone to high volatility or influenced by external macroeconomic and geopolitical factors not captured in the price history alone \citep{Khilji1993StockReturns,Sohail2009MacroeconomicPakistan}.

Several hypotheses can explain these differential outcomes:
\begin{itemize}
    \item \textbf{Hypothesis 1: Stability Enables Pattern Learning} — Stocks in stable, liquid sectors (e.g., consumer goods, utilities) exhibit relatively consistent trading patterns and lower noise levels in price movements. This allows the LSTM network to effectively learn and extrapolate temporal dependencies from historical data, leading to higher predictive accuracy.
    
    \item \textbf{Hypothesis 2: Volatility Disrupts Temporal Coherence} — Highly volatile stocks often experience sharp, non-recurring price swings driven by speculation, short-term sentiment, or sudden news. These abrupt changes disrupt the sequential structure that LSTMs rely on, making it difficult to distinguish between meaningful trends and random fluctuations.
    
    \item \textbf{Hypothesis 3: Absence of Exogenous Variables Limits Predictive Power} — Many stocks in emerging markets like Pakistan are significantly influenced by external factors such as political instability, regulatory changes, currency fluctuations, or global commodity prices. Since the model relies solely on historical price data, it lacks critical contextual information, resulting in poor performance during periods dominated by such extrinsic shocks \citep{Sohail2009MacroeconomicPakistan,Oukhouya2025MoroccoPrediction}.
    
    \item \textbf{Hypothesis 4: Market Efficiency Varies by Sector} — More liquid sectors may exhibit semi-strong form efficiency, where past prices partially reflect available information, enabling data-driven models like LSTM to extract useful signals. In contrast, less liquid or speculative sectors may behave more randomly, limiting the effectiveness of purely technical forecasting approaches \citep{Khan2006WeakForm}.
\end{itemize}
    
These findings suggest that while LSTMs are powerful tools for time-series forecasting, their success in financial markets depends heavily on the underlying structural and behavioral characteristics of the target asset. Future work could explore hybrid models that incorporate exogenous variables (e.g., news sentiment, macroeconomic indicators) to improve robustness in volatile or information-sensitive environments \citep{Ziolkowski2025WaveletLSTM,Buczynski2023DeepLearningReview}.

\subsection{Implications and Limitations}
The findings are significant for investors and analysts operating in emerging markets, providing a framework for employing deep learning models even with limited data. The model's sector-dependent performance suggests it is better suited for informing medium-term investment strategies rather than high-frequency trading \citep{Ofonedu2022StockAnalysis,Mehdi2024RiskPrediction}.

Several limitations must be acknowledged. The model assumes historical patterns are indicative of future performance, an assumption that fails during black swan events or structural market shifts. Data quality and availability for some stocks constrained model accuracy. Crucially, the model does not incorporate external variables like news sentiment, political events, or global commodity price shocks, which are significant drivers of price movements in markets like Pakistan \citep{Khilji1993StockReturns,Sohail2009MacroeconomicPakistan}.

\subsection{Future Work}
Future research should focus on integrating alternative data sources (e.g., news headlines, social media sentiment, macroeconomic indicators) into the model architecture. Hybrid modeling approaches, combining LSTMs with other techniques like GARCH for volatility modeling or Attention Mechanisms, could improve performance for volatile stocks. Finally, a rigorous analysis of practical trading strategies based on these predictions would solidify the real-world applicability of the model \citep{Zaheer2023MultiParameter,Ziolkowski2025WaveletLSTM}.

\section*{Acknowledgements}
We are grateful to the Department of Economics, LUMS for the cooperation rendered in the collection of raw data for stocks and its validation.

\end{document}